\documentstyle[twoside,prl,aps]{revtex}

\begin{document}

\input epsf
\draft

\title{Indication of Anisotropy in Electromagnetic Propagation over
Cosmological Distances}

\author{Borge Nodland}

\address{Department of Physics and Astronomy, and Rochester Theory
Center for Optical Science and Engineering, University of Rochester,
Rochester, NY 14627}

\author{John P. Ralston}

\address{Department of Physics and Astronomy, and Kansas Institute for
Theoretical and Computational Science, University of Kansas, Lawrence,
KS 66044}

\date{Published in Physical Review Letters {\bf 78,} 3043 (1997)}

\maketitle

\begin{abstract}
We report a systematic rotation of the plane of polarization of
electromagnetic radiation propagating over cosmological distances.  The
effect is extracted independently from Faraday rotation, and found to
be correlated with the angular positions and distances to the sources.
Monte Carlo analysis yields probabilistic P-values of order $10^{-3}$
for this to occur as a fluctuation. A fit yields a birefringence scale
of order $10^{25} \frac{h_0}{h} \mbox{m}$. Dependence on redshift $z$
rules out a local effect. Barring hidden systematic bias in the data,
the correlation indicates a new cosmological effect.
\end{abstract}              


Polarized electromagnetic radiation propagating across the universe has
its plane of polarization rotated by the Faraday effect \cite{one}. We
report findings of an additional rotation, remaining after Faraday
rotation is extracted, which may represent evidence for cosmological
anisotropy on a vast scale.

We examined experimental data \cite{one} on polarized radiation emitted
by distant radio galaxies. The residual rotation is found to follow a
dipole rule, depending on the angle $\gamma$ between the propagation
wavevector $\vec k$ of the radiation and a unit vector $\vec s$. The
rotation is linear in the distance $r$ to the galaxy source; in sum,
the rotation is proportional to $r \cos(\gamma)$. This effect can not
be explained by uncertainties in subtracting Faraday rotation. We focus
on a statistical analysis of the correlation, but we have also made
considerable effort to explain it in a conventional way. Unless the
effect is due to systematic bias in the data, it seems impossible to
reconcile it with conventional physics.

Some history is useful. In 1950, Alfven and Herlofson \cite{one}
predicted that synchrotron radiation would be emitted from galaxies,
with polarization perpendicular to the source magnetic field. By the
mid 1960's, data began to accumulate on the polarization of radio waves
that had traveled over cosmological distances \cite{one}. The
observables include the redshift $z$  of the galaxy source, an angle
$\psi$ labeling the orientation of the galaxy major axis, the percent
magnitude of polarization $p$, and angles $\theta(\lambda)$ labeling
the orientation of the plane of polarization of radio waves of
wavelength $\lambda$. Experimental fits \cite{one} show that the angle
$\theta(\lambda)$ for a galaxy is given by $\theta(\lambda) =
\alpha\lambda^2 + \chi$. Fits to the linear dependence of the angle
$\theta$ on $\lambda^2$ verify the presence of Faraday rotation. The
fitting parameter $\alpha$, called the Faraday rotation measure,
depends upon the magnetic field and the electron density along the line
of sight \cite{one}. Conventional Faraday rotation does not account for
angle $\chi$, the orientation of the polarization plane after Faraday
rotation is taken out; $\chi$ is central to our analysis.

On symmetry grounds, $\chi$ would be expected to approximately align
with the major axis angle $\psi$ of a galaxy. This expectation has
consistently been at odds with the data. Gardner and Whiteoak
\cite{one}  proposed a ``two--population'' hypothesis, at first on the
basis of 16 sources, with some sources emitting at $\chi-\psi =
0^\circ$, and others at $\chi-\psi = 90^\circ$.  Clarke {\it et al.}
\cite{one} found a subset of galaxies supporting the two--population
idea, but by making a severe cut consisting of galaxies with quite
strong polarization, which eliminated most of the data ($p>11$, leaving
47 of 160 galaxies.) The group selected as perpendicular emitters was
found to be distant (high luminosity), while the parallel group was
found to be near (low luminosity).  The full data set (no strong
polarization cuts) does not convincingly support multi--populations,
and the statistical significance of conclusions is not given.  The
reader is warned of inconsistencies in the statistics: in Clarke {\it
et al.}\cite{one}, the quantity $\Delta = |\chi - \psi |$ is defined to
be the statistic, but actually applied when  $|\chi - \psi | < \pi/2$,
while $\Delta =\pi - |\chi - \psi |$ is used otherwise. Carroll {\it et
al.} \cite{eight} use $\chi- \psi$ as a measure of residual rotation, a
definition which neglects the other possible rotation
$\chi-\psi\pm\pi$. (A few errors in this paper's transcription of the
data have been corrected.)

Birch in 1982 observed a dipole rule correlation of polarization angles
and source location angles relative to an axis he fit from the data
\cite{two}. Birch used the acute angle between $\chi$ and $\psi$ in a
limited sample of data. The acute angle is an improper statistic for
the observables, which are not vectors, but planes. Kendall and Young
(KY) confirmed Birch's conclusions \cite{two} using a proper
(projective) statistic. Bietenholz and Kronberg (BK) also confirmed the
same correlation with a different analysis \cite{two}. However,
introducing a larger sample of data with sources for which redshifts
were not known, BK then found no correlation of Birch's type. Although
the subject died out after BK's negative conclusion, the history is
relevant inasmuch as the redshift is found to play a role.

None of these studies addressed a correlation going like $r
\cos(\gamma)$, the $\cos(\gamma)$ being the lowest order anisotropic
effect that might be observed, and the factor $r$ representing the
generic dependence of birefringence on distance. The length $r$ must
be measured in a basic unit denoted $\Lambda_s$. An ansatz for the
residual rotation angle $\beta$ is

\begin{equation}
\beta=\frac{1}{2}\Lambda_s^{-1} r \cos(\gamma).
\label{eq1}
\end{equation}

The astronomical literature has several, sometimes inconsistent
measures of angle differences. The angles $\chi$ and $\psi$ do not
label vectors, but plane orientations; they are defined only up to
multiples of $\pi$ (not 2$\pi$). Analysis should retain information on
the sign of the differences of $\chi$ and $\psi$ which probes the sense
(clockwise or counterclockwise) of the rotation. We introduce two
functions $\beta^+$ and $\beta^-$, given by

\begin{equation}
\beta^+ = 
\left\{ \begin{array}{ll}
\chi-\psi & \mbox{if $\chi-\psi\geq 0$}\\
\chi-\psi+\pi & \mbox{if $\chi-\psi < 0$}
\end{array}
\right.
\;\;\;\;\;
\beta^- = 
\left\{ \begin{array}{ll}
\chi-\psi & \mbox{if $\chi-\psi < 0$}\\
\chi-\psi-\pi & \mbox{if $\chi-\psi\geq 0.$}
\end{array}
\right.
\label{eq2}
\end{equation}

 From Eq. (\ref{eq1}), the rotation $\beta$ is either positive or
negative depending on the angle $\gamma$. We therefore assign a
rotation $\beta=\beta^+_i$  to a galaxy ($i$) if $\cos(\gamma_i) \geq
0$, and a rotation $\beta=\beta^-_i$ if $\cos(\gamma_i) < 0$.  This
assignment necessarily introduces correlations because two quadrants of
the data plane of $\beta_i$ and $r_i\cos(\gamma_i )$ are excluded, a
point which we discuss momentarily.

By definition, $\beta$ lies in the interval $[-\pi,\pi]$. If the
residual rotation magnitude is bigger than $\pi$ for certain sources,
then that data would be scrambled and yield low correlations.
Fluctuations in the initial orientation will contribute noise to the
analysis. Our data consists of the most complete set we have found of
160 sources in \cite{one} which includes polarization as well as
position information. The radio frequency varies, typically spanning a
1--3 GHz range. Measurement uncertainties, less than $5^\circ$ for
$\psi$ and typically $5^\circ$ for $\chi$, were not reported on a
point--by--point basis. Since the existing Faraday fits to $\psi=
\alpha\lambda^2+\chi$ are done using several data points, the errors on
$\chi$ should be much smaller than a few degrees. Galaxy position
coordinates \cite{one} are given in terms of distance ($r$), right
ascension (R.A.), and declination (Decl.). The majority of the data
comes from the northern sky, with visual magnitude of the galaxies
between 8 and 23. We used the $\Omega =1$ (critical average mass
density) relation for the distance $r$ traveled by light as a function
of redshift $z$, namely $r= 10^{10} \mbox{(lightyears)} \lbrack
1-(1+z)^{-3/2}\rbrack \frac{h_0}{h}$, where $h_0=\frac{2}{3} 10^{-10}
\mbox{(years)}^{-1}$ and $h$ is the Hubble constant.

To determine the validity of Eq. (\ref{eq1}), we computed the linear
correlation coefficient $R_{\text{data}}$ for the 160 points in the
galaxy data set of [$r_i\cos(\gamma_i) , \beta_i$] for trial values of
the $\vec s$--direction $(\text{Decl.},\text{R.A.})_s$ sweeping out all
directions.  For a general set $(x_i,y_i)$ of $N$ datapoints, $R$ is
defined as $R=(N\Sigma x_iy_i- \Sigma x_i\Sigma y_i)/ \{[N\Sigma
x^2_i-(\Sigma x_i)^2]^\frac{1}{2} [N\Sigma y^2_i-(\Sigma
y_i)^2]^\frac{1}{2}\}$.  For each trial direction of $\vec s$, we also
computed 1000 correlation coefficients $R_{\text{rand}}$ from 1000
copies of the original data set ($160 \times 1000 \times 2 = 320,000$
false data points), where the copies had rotations $\beta_i$ obtained
from substituting random $\psi_i$ and $\chi_i$ into (\ref{eq2}). The
positional part of the data, $r_i \cos(\gamma_i)$, was not randomized.
In randomizing only the polarizations and major axis orientations, we
created a sample with the same spatial distribution of points as the
data itself, thus allowing a scan of the polarization data without
prejudice caused by the $\beta^{\pm}$ assignment, or the spatial
non--uniformity of the data's distribution.

\begin{figure}
\centerline{
\epsfxsize= 0.5 \textwidth
\epsfbox[150 230 500 670]{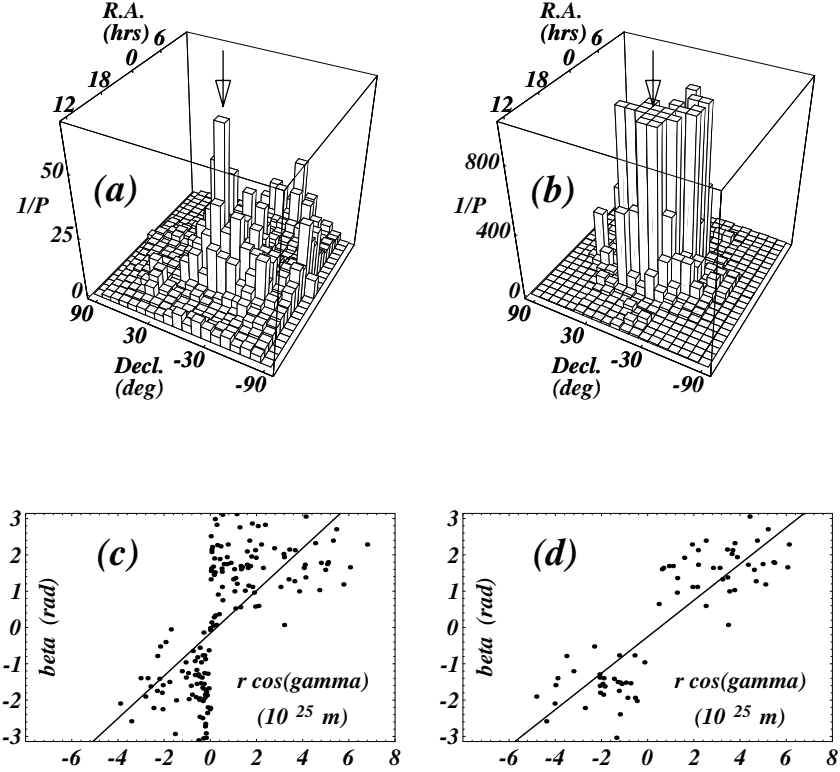} }
\caption{
Inverse P--values versus anisotropy direction $\vec s$, given in terms
of its declination and right ascension. $P$ is the fraction of galaxy
sets with randomized major axis $(\psi_{\text{rand}})$ and polarization
$(\chi_{\text{rand}})$ angles that yielded a linear correlation
coefficient $R_{\text{rand}}$ of the set [$r_i\cos\gamma_i,
\beta_i(\psi_{\text{rand}},\chi_{\text{rand}})$] greater than or equal
to the linear correlation coefficient $R_{\text{data}}$ of the original
set [$r_i \cos\gamma_i,\beta_i(\psi,\chi)$]. (a) All data (160
galaxies).  (b) All data with redshift $z\geq 0.3$ (71 galaxies). The
figures support an anisotropy direction in the region $\vec{s}^* =
(\text{Decl.}, \text{R.A.})^*_s =(0^\circ \pm 20^\circ, 21 \mbox{hrs} \pm
2\mbox{hrs})$. (c) Scatter plot of $\beta$ versus $r\cos\gamma$ for the
$\vec s$--direction yielding the highest peak (arrow) of (a) for all
data. (d) Scatter plot of $\beta$ versus $r\cos\gamma$ for the $\vec
s$--direction yielding a typical central peak (arrow) of (b) for the
cut $z\geq 0.3$.}
\label{fig1}
\end{figure}

For each trial $\vec s$--direction, we then compared the
$R_{\text{data}}$ from the data set with the $R_{\text{rand}}$
distribution, by computing the fraction $P$ of computations with
$R_{\text{rand}} \geq R_{\text{data}}$.  In statistics, $P$ is called a
P--value; the interpretation as a ``probability'' depends on various
assumptions and details of terminology. A plot of $P^{-1}$ is shown in
Fig. \ref{fig1}(a). The result is stable and scaled properly as we
increased the number of independent trial orientations of $\vec s$; the
figure shows our finest resolution of 410 bins covering the entire
celestial sphere with bins of average solid angle of about $\pi/100$.
There is a clear excess in the $P^{-1}$--plot in the region $\vec{s} =
(\text{Decl.}, \text{R.A.}) = (-10^\circ \pm 20^\circ, 20 \mbox{hrs}
\pm 2 \mbox{hrs})$.

To explore this, we cut the data to $z\geq 0.3$, roughly the most
distant half of the sample (71 galaxies), to improve the experimental
``lever arm.'' The correlation for the $z\geq 0.3$ set is much more
dramatic; we see in Fig. \ref{fig1}(b) a well--connected cluster of
more than twenty-one peaks in the region $\vec{s}^* = (\text{Decl.},
\text{R.A.})^*_s = (0^\circ \pm 20^\circ, 21 \mbox{hrs} \pm 2
\mbox{hrs})$ with a P--value lower than we can resolve ($P\leq
10^{-3}$).  [Several of the $\vec s$-directions displayed in Fig.
\ref{fig1}(b) had no Monte Carlo events with $R_{\text{rand}} \geq
R_{\text{data}}$ in the 1000-trial runs.  $P=10^{-3}$ was assigned to
these directions.] We call this procedure 1.

For the $\vec s$--direction with highest $P^{-1}$--value of the full
data set, the distribution of $R_{\text{rand}}$ is a Gaussian, centered
at $\mu=0.60$ with a standard deviation $\sigma=0.032$, with
$R_{\text{data}}=0.66=\mu+1.88 \sigma$. In contrast, in a typical
``off--axis'' direction $(60^\circ, 12 \mbox{hrs})$, the distribution
is given by $\mu=0.47$ and $\sigma=0.04$, with
$R_{\text{data}}=0.48=\mu+0.25 \sigma$.  Returning to the strongly
correlated data set, with $z\geq 0.3$, and for an $\vec{s}$--direction
yielding a high $P^{-1}$--value, a typical $\mu=0.76$ and
$\sigma=0.027$, with $R_{\text{data}}=0.86=\mu+3.7\sigma$.
Distributions with long tails were not seen. The spatial distribution
of galaxies in the sample is quite non--uniform, so that the
populations assigned to $\beta^\pm$, and the $\mu$ and $\sigma$ values,
depend on the trial $\vec s$. The P--values are therefore much more
meaningful than the correlation coefficients $R$ themselves. We include
a few scatter plots [e.g. Figs. \ref{fig1} (c), (d)], but warn that
visual inspection is not very reliable.  For the set of strongly
polarized sources with polarization $p\geq 5$ (116 galaxies), the plot
of $P^{-1}$ versus $\vec s$ is almost identical to Fig. \ref{fig1}(a).
We also studied the dependence in shells of $z_{min}<z<z_{max}$
containing 20 galaxies, which we considered a minimum number for a
sensible analysis. Only in the region $0<z<0.2$ was $P\geq 10^{-2}$;
bin by bin, $P<10^{-3}$ in the region $\vec s$=$(0^\circ \pm 20^\circ,
21 \mbox{hrs} \pm 2 \mbox{hrs})$ thereafter.

These studies use the correlation coefficient $R$ appropriate to test
unconstrained linear fits of the form $\beta=\frac{1}{2}\Lambda_s^{- 1}
r\cos(\gamma) + \delta$, where $\Lambda_s$ and $\delta$ are free
parameters. We found that $\delta$ is consistent with zero $(\delta
\leq 2^\circ$, with $\delta$ typically $\approx 1^\circ$) in the region
of good correlation ($z\geq 0.3,\; P^{-1}\geq 1000$ or $z\geq 0,\;
P^{-1}\geq 25$). We also repeated the entire study using $R=\sum x_i
y_i / (\sum x_i^2 \sum y_i^2)^{1/2}$, which tests the hypothesis of
linear correlation with the constraint $\delta=0$, without finding
significant changes.

Another study (procedure 2) used a different order. For each random
data set (again, galaxy positions were not randomized, as explained
above), we varied $\vec s$ over the sphere (410 directions) to maximize
$R_{\text{rand}}$. This ``largest--$R_{\text{rand}}$'' value was then
recorded. A new random set was then generated, producing another
``largest--$R_{\text{rand}}$.'' This calculation was repeated more than
1000 times, to create a set of largest $R_{\text{rand}}$'s.  This
procedure was motivated by the fact that there is an increased
probability in procedure 1 of obtaining a fit of $\vec s$ to the data
due to the two degrees of freedom of $\vec s$. The crucial test is for
the far--half data set with $z\geq 0.3$, which had a P--value of order
$10^{-3}$ (in procedure 1). For the far--half sample with $z\geq 0.3$,
we found that the fraction  of the largest--$R_{\text{rand}}$'s that
exceeded $R_{\text{data} (z \geq 0.3, \vec{s}=\vec{s}^*)}$ was less than
0.006. The distribution of largest--$R_{\text{rand}}$'s was
characterized by a $\sigma$ and $\mu$ such that $R_{\text{data}}=\mu+2.8
\sigma$. In contrast, for the closest half of the data, $z<0.3$, the
fraction of the largest $R_{\text{rand}}$ exceeding $R_{\text{data}
(z<0.3,\vec{s}=\vec{s}^*)}$ was 0.86, confirming that the effect
``turns on'' only for the most distant half of the galaxies.  We found
no instances of largest--$R_{\text{rand}}$'s for the sample with
$z<0.3$ exceeding $R_{\text{data} (z \geq 0.3, \vec{s}=\vec{s}^*)}$. These
results provide a more stringent test and corroborate the conclusion of
procedure 1.  (The authors welcome requests for additional
information.)

The average (procedure 1) best fit value is $\Lambda_s= (1.1 \pm 0.08)
10^{25} \frac{h_0}{h} \mbox{m}$ for a $\vec s$--direction of
$(\text{Decl.}, \text{R.A.})^*_s = (0^\circ \pm 20^\circ, 21 \mbox{hrs}
\pm 2 \mbox{hrs})$ for the data with $z\geq 0.3$. For the full  data
set of all 160 data points, we find $\Lambda_s = ( 0.89 \pm 0.12)
10^{25} \frac{h_0}{h} \mbox{m}$ for $(\text{Decl.}, \text{R.A.}) =
(-10^\circ \pm 20^\circ, 20 \mbox{hrs} \pm 2 \mbox{hrs})$. The scale
$\Lambda_s$, approaching a billion parsecs, is approximately an order
of magnitude larger than the largest scales observed in galaxy
correlations. (Errors in $\Lambda_s$ are the usual 1 sigma variation of
uncorrelated analysis and do not refer to probabilities.) The direction
$\vec{s}^*$ appears unremarkable; although vaguely toward the galaxy
center, the cluster in Fig.  \ref{fig1}(b) is separated from the galaxy
center by $30^\circ$, about 30 times the apparent size of the core
region of the Milky Way.

As a consistency check, we have separately investigated whether there
are strong correlations in $\beta=\Lambda_s^{-1} r$, and
$\beta=\beta_0\cos(\gamma)$. We find nothing significant in the first
case, or for the second case over the full data set. The set $z\geq
0.3$ does produce correlations with $\beta_0\cos(\gamma)$ which we
cannot readily distinguish from Eq. (\ref{eq1}), since the $r$--values
do not vary enough.

As for conventional physics, the effect observed is not explained by
variations on Faraday physics. While fits to Faraday rotation (linear
in $\lambda^2$) represent a model and an approximation, the ratios of
the radio frequency to cyclotron and plasma frequencies are such that
the approximation is thought to have exceedingly small corrections.
Observers also make corrections for systematic errors and take into
account the effects of the Earth's ionosphere. Consulting with the
original observers \cite{three} does not yield any suggestions for bias
that would imitate the signal we observe. We have nevertheless
questioned whether a local effect of the galaxy, via some unanticipated
conventional physics, might account for our correlation. The fact that
the correlation is seen for $z \geq 0.3$, but not $z< 0.3$, rules out a
local effect. (Several observational groups contributed to the $z\geq
0.3$ data, closing the loophole that one particular analysis might
contain bias, as best we can determine.)  Again, strong fields at the
source might generate unexpected initial polarization orientations, or
upset the Faraday--based fits, and this could plausibly depend on $z$.
But since the correlation is observed in $\cos(\gamma)$, any
population--based explanation requires an unnatural, if not impossible,
conspiracy between distant sources at widely separated zenith angles.
One is left, then, with the option of unknown systematic bias for the
large--$z$ set, or accepting the possibility that the correlation is a
real physical effect. For the latter, one must arbitrarily invoke
coherence on outrageously vast distances, perhaps organized by
electromagnetic or other interactions in the early universe, or
contemplate new physics.

If we take the data at face value as indicating a fundamental feature
of electrodynamics, gauge invariance severely limits the possible
couplings of the vector potential $A^\mu$ and the electromagnetic field
strength tensor $F^{\mu\nu}= \partial^\mu A^\nu - \partial^\nu A^\mu$
to any background vector $s^\mu$. The unique derivative expansion
(units are $\hbar = c =1$) for terms in the effective action $S_{eff}$
is

\begin{equation}
S_{eff} =\int d^4 x\biggl(-{1\over 4}F_{\mu\nu}F^{\mu\nu}+{1\over 
4} \Lambda^{-1}_s\varepsilon^{\mu\nu\alpha\beta}F_{\mu\nu}A_\alpha
s_\beta\biggr),
\label{eq3}
\end{equation}
suppressing higher derivative terms which would contribute to
short--distance effects. The dispersion relation for this theory at
lowest order in $\Lambda_s^{-1}$ is $k_\pm = \omega\pm
\frac{1}{2}\Lambda_s^{-1} \cos(\gamma)$, where $s^\mu=(0,\vec s)$ in
the coordinate system where $\omega$ and $k$ are measured. Rotation of
the plane of polarization comes from differences in propagation speed
between the two modes; the difference $\frac{1}{2} (k_{+} - k_{-}) = d
\beta/dr$ is a measure of the polarization plane rotation $\beta$ per
unit path length $r$, yielding a rotation $\beta$ coinciding with Eq.
(\ref{eq1}).

The interpretation of the parameters $s^\mu$ depends on their physical
origin and their transformation properties. If $s^\mu$ is odd under
time reversal, and its space part is a pseudovector under parity, then
Eq. (\ref{eq3}) preserves these symmetries separately. If $s^\mu$ is
also spacelike, as we have assumed, one might associate the vector
$\vec{s}$ with an intrinsic ``spin axis'' of an anisotropic universe.
Terms of order $s^0$ are dropped consistently if our reference frame
coincides well enough with a rest frame of $s^\mu$; $s^0$ can also be
added as a separate parameter. Because the new terms have one less
derivative than the standard ones, these terms have no effect (by power
counting) on high energy questions such as renormalizability. The scale
$\Lambda_s$ represents a fundamental length scale in the modified
electrodynamics due to Eq. (\ref{eq3}), and should not be confused
with a ``photon mass'', which violates gauge invariance. By converting
the length $\Lambda_s$ to a mass scale $\mu c^2= \Lambda_s^{-1} \hbar
c$, a value of $\mu = 10^{-32} \mbox{eV}/c^2$ is found, which is
$10^{14}$ times smaller than the photon mass limit of Chibisov, and
$10^{17}$ times smaller than Goldhaber and Nieto's \cite{four}.

Ni \cite{five} obtained (\ref{eq3}) from covariance arguments;
independently it was found from quantum adiabatic arguments (Ralston
\cite{five}). The latter was our initial motivation, used to predict
the correlation (\ref{eq1}), which led to this investigation. The
curious history came to light much later. If $s^\mu$ is treated as
dependent on $x$, then it must be a gradient: $\Lambda_s^{-1} s^{\mu} =
\frac{g}{8}\partial^{\mu}\phi$, where $g$ is a coupling. The theory is
then related to axions \cite{seven} or similar pseudoscalar fields
$\phi$ with coupling $g$. If a new field is proposed, there should be
further observational consequences. Our study of a spacelike $s^\mu$
is unconventional compared to other work \cite{eight}, but there is
every reason to think that the observed correlation can be consistent
with axion-type domain walls \cite{seven}, or other condensate
structures. More data on the observables is needed, especially from the
southern sky. The crucial issues of conventional explanations and
experimental systematics merit scrutiny from a broad community. From a
scientific standpoint, we report what we find, given the data that
exist. We find that the data contain a correlation indicating
cosmological anisotropy in electromagnetic propagation. Further study
may be able to determine whether (\ref{eq3}) or its counterparts
invoking new fields might be a valid description of electromagnetism on
the largest scale.

B. Anthony-Twarog, K. Ashman, C. Bird, H. Rubinstein, B. Cox, and an
anonymous referee made helpful suggestions. DOE Grant Number
DE-FG02-85ER40214, NSF Grant Number PHY94-15583, and the $K^{*}STAR$
program provided support.

\end{document}